\documentclass[12pt]{article}

\usepackage{graphicx}
\usepackage{amsmath,amssymb}
\usepackage{euscript}
\usepackage{a4wide}
\usepackage{booktabs,dcolumn,multirow,colortbl}
\usepackage[pdftex,table]{xcolor}
\usepackage{units}
\usepackage{slashed}
\usepackage{cite}

\begin{document}
\begin{titlepage}
\begin{flushright}
SI-HEP-2008-07 \\[0.2cm]
CERN-PH-TH/2008-093 \\[0.2cm]
May 7, 2008
\end{flushright}

\vspace{1.2cm}
\begin{center}
\Large\bf\boldmath
Role of ``Intrinsic Charm''  \\ in Semi-Leptonic $B$\/-Meson Decays
\unboldmath
 \end{center}

\vspace{0.5cm}
\begin{center}
{\sc C.~Breidenbach$\,^a$, T.~Feldmann$\,^a$, T.~Mannel$\,^{a,b}$ and S.~Turczyk$\,^a$} 

\vspace{1.4em}

${}^a$ {\sf Theoretische Physik 1, Fachbereich Physik,
Universit\"at Siegen,\\ D-57068 Siegen, Germany.}
\vspace{0.7em}

${}^b$ {\sf CERN, Department of Physics, Theory Unit, \\ CH-1211 Geneva 23,
Switzerland.}

\end{center}

\vspace{3em}
\begin{abstract}
\vspace{0.2cm}\noindent
We discuss the role of so-called ``intrinsic-charm'' operators 
in semi-leptonic $B$\/-meson decays, which appear first at 
order $1/m_b^3$ in the heavy quark expansion.
We show by explicit calculation that -- at scales $\mu \leq m_c$ --
the contributions from ``intrinsic-charm'' effects can be absorbed 
into short-distance coefficient functions multiplying, for instance, the Darwin term. 
Then, the only remnant of ``intrinsic charm'' are
logarithms of the form $\ln (m_c^2 / m_b^2)$, which can be resummed
by using renormalization-group techniques. As long as the dynamics at the charm-quark
scale is perturbative, $\alpha_s(m_c) \ll 1$, this implies that  
no additional non-perturbative matrix elements aside from the Darwin 
and the spin-orbit term have to be introduced at order $1/m_b^3$. 
Hence, no sources for additional hadronic uncertainties have to be taken into account. 
Similar arguments may be made for higher orders in the $1/m_b$ expansion.

\end{abstract}

\end{titlepage}

\section{Introduction}
The heavy quark expansion (HQE) has turned out to be a valuable tool for precision calculations of 
heavy hadron decays \cite{Chay:1990da,Bigi:1993fe,Manohar:1993qn,Mannel:1993su}. 
In particular, due to the HQE for semi-leptonic decays, where the $b \to c$ transition is described 
in the framework of a standard local OPE, the relative uncertainty in 
the CKM matrix element  $|V_{cb}|$ could be reduced to a level below 2\% 
\cite{Benson:2003kp,Gambino:2004qm,Bauer:2002sh,Hoang:1998hm}.

The expansion in inverse powers of the 
heavy quark mass $m_b$ can be set up for both, the lepton-energy spectrum
as well as for the total decay rate.
The non-perturbative input, entering the theoretical description,
is given by forward matrix elements of local operators in the OPE. 
The leading term represents the partonic rate and does not contain 
any unknown hadronic matrix element.
The perturbative corrections to the partonic rate have 
been calculated  to order $\alpha_s^2$, recently \cite{Melnikov:2008qs,Pak:2008qt}. 
Terms of order $1/m_b$
vanish due to heavy quark symmetries. 
At order $1/m_b^2$, two hadronic parameters $\mu_\pi^2$ and $\mu_G^2$ appear,
which can be interpreted  as the kinetic energy and the chromomagnetic moment 
of the heavy quark inside the heavy hadron. 
The short-distance contribution to the coeffcient of $\mu_\pi^2$ is 
known to order $\alpha_s$ \cite{Becher:2007tk}, while that of $\mu_G^2$ is known at tree level.
The dimension-6 operators at order $1/m_b^3$ define two additional parameters,
which correspond to the Darwin term $\rho_D^3$ and the spin-orbit term $\rho_{\rm LS}^3$,
known from the usual non-relativistic reduction of the Dirac equation. 
The coefficients at that order are only known at tree level, so far. 
The terms at order  $1/m_b^4$ have also been classified,
and introduce five new hadronic parameters \cite{Dassinger:2006md}.

It has also been pointed out that at order
$1/m_b^3$ a dimension-6 operator appears,  whose matrix element 
could be interpreted as the ``intrinsic-charm'' 
content of the $B$\/-meson \cite{Bigi:2005bh,Brodsky:2001yt}. 
An order-of-magnitude estimate for the effect has been 
given in \cite{Bigi:2005bh}, and the additional uncertainty 
from the poor knowledge of these matrix elements 
has been included in the error budget for $|V_{cb}|$ \cite{Buchmuller:2005zv}. 
However, as we are going to show in this paper, the inclusion of an
``intrinsic-charm'' contribution requires a proper definition of the short-distance
functions appearing in the lepton-energy spectrum, since the ``intrinsic-charm''
operators and, for instance, the Darwin term mix under renormalization.
As long as the strong dynamics at the charm-mass scale is treated perturbatively,
the effect of ``intrinsic charm'' can entirely be absorbed into short-distance
coefficients defined at a low hadronic input scale, 
and the non-analytic dependence on the charm-quark mass can
be resummed by standard renormalization-group techniques, extending
the results in \cite{Bauer:1996ma}. 
In this case, no additional hadronic uncertainty due to ``intrinsic charm'' 
has to be included.
On the other hand, treating the charm-quark as non-perturbative, the
hadronic matrix elements of intrinsic-charm operators would remain as unknown parameters. In this case, however, the charm-quark dependent terms in the standard expressions for the lepton-energy spectrum and the total rate have to be modified accordingly, in order to avoid double counting.

In this paper, we are going to present a systematic study
of how ``intrinsic-charm'' effects will enter the theoretical
expressions for the lepton-energy spectrum, depending on the
treatment of the charm-quark mass scale, with particular
emphasis on the mixing of the ``intrinsic-charm'' operators
into the Darwin term.


\section{Calculation of the Charm Contribution} 

Starting point for the calculation of inclusive rates within the OPE
is the hadronic tensor $W_{\mu\nu}$
as it appears in the differential rate for  
$b \rightarrow c \ell \bar{\nu}_\ell$ transitions,
\begin{align}
d\Gamma &= 16 \pi \, G_F^2 \, |V_{cb}|^2 \, W_{\mu\nu} \, L^{\mu\nu}
 \, d\phi \,.
\end{align}
Here $d\phi$ denotes the invariant phase space for the lepton-neutrino pair, 
and the leptonic tensor is given by
\begin{align}
 L^{\mu\nu} &= 2 \left( p^\mu_e p^\nu_{\nu_e} + p^\nu_e p^\mu_{\nu_e} 
 - g^{\mu\nu} \, p_e \cdot p_{\nu_e} - i \epsilon^{\mu\nu\alpha\beta} \, p_e{}_\alpha\,
  p_{\nu_e}{}_\beta \right) \,,
\end{align} 
where $\epsilon^{0123}=-\epsilon_{0123}=+1$.
Using translational invariance, the hadronic tensor may 
be cast into the form 
\begin{equation}
2M_B W_{\mu \nu} = \int d^4 x \,  e^{i(m_b v-q)x} \, 
\langle \bar B(p) |\bar{b}_v (x) \, \gamma_\nu P_L \, c(x)
\ \bar{c}(0) \, \gamma_\mu  P_L \, b_v(0)  | \bar B(p) \rangle \,  ,
\end{equation} 
where $P_L = (1-\gamma_5)/2$ projects onto left-handed fields,
$v^\mu = p^\mu/M_B$ is the velocity of the decaying $\bar B$\/-meson, and
$b_v(x)$ denotes the heavy $b$\/-quark field with the phase 
$e^{-i m_b v \cdot x}$ factored out. Performing the OPE for this matrix element,
the product of the two $b \to c$ currents is matched onto a set of local
operators at scales $\mu$ of the order of the $b$\/-quark mass $m_b$.
Now, as far as the charm-quark mass is concerned, one may take different points of view
\cite{Mannel:1994pm}:
\begin{enumerate}

\item One may assume that $m_b \sim m_c \gg \Lambda_{\rm QCD}$,
 which means that the short-distance matching coefficients
 and the phase space integrals are functions of the fixed ratio $\rho=m_c^2 / m_b^2$.
 In other words, one integrates out (hard) quantum fluctuations with virtualities of
 order $m_{b,c}^2$ and is left with light-degrees of freedom: light quarks and gluons,
 together with the quasi-static $b$\/-quark field in HQET.
 In a standard renormalization scheme like $\overline{\rm MS}$, operators with
 charm fields do not appear at scales $\mu < m_c$. 
 More precisely, such operators would correspond to quasi-static charm quarks,
 which cannot contribute to the considered matrix elements, $\langle \bar B| \bar b_v
   \ldots c_{\rm static} \, \bar c_{\rm static} \ldots b_v|\bar B \rangle \equiv 0$, because of energy conservation, $m_b + 2 m_c + \Delta E_{\rm soft} > m_B$.
This is in fact the point of view  that is usually considered in the precision determination of $|V_{cb}|$.

\item One may consider the power counting $ m_b \gg m_c \gg \Lambda_{\rm QCD}$,
  and integrate out hard $b$\/-quark fluctuations at a different scale 
  than the charm quark. In this case, for the first matching at the high scale
  $\mu_h \sim m_b$ one still has to keep the charm quark dynamical, and the
  corresponding ``intrinsic-charm'' operators appear in the OPE.\footnote{More precisely,
  the ``intrinsic-charm'' operators correspond to local operators for semi-hard
  charm quarks,
  i.e.\ quarks with all momentum components of order $m_c$. This is to be distinguished
  from the hard-collinear (jet) modes for  the charm-quark which appear in
  non-local operators describing 
  the shape-function region \cite{Mannel:2004as,Boos:2005by} for $b \to c\ell\nu$.}
  The renormalization-group for these operators can be used
  to scale down to the semi-hard scale $\mu_{sh} \sim m_c$, where the
  charm-quark is finally integrated out. As before, the ``intrinsic-charm'' operators
  then match onto local operators built from light fields.
  Obviously, the main difference compared to case~1 is, that
  logarithmic terms $\ln (m_c/m_b)$ can be resummed into short-distance coefficient
  functions \cite{Bauer:1996ma}, while the analytic terms should be
  expanded in powers of $m_c/m_b \sim \sqrt{\Lambda_{\rm QCD}/m_b} \sim 0.3$.

\item Finally, one may assume that $ m_b \gg m_c \gtrsim \Lambda_{\rm QCD}$.
      In this case, one cannot integrate out the charm-quark effects 
      perturbatively, and is thus left with genuine intrinsic-charm operators,
      whose hadronic matrix elements have to be defined at a sufficiently
      high scale $\mu_0$, satisfying $m_b \geq \mu_0 \gg m_c$. 
      Notice that the matrix elements of the intrinsic-charm operators
      contain the non-analytic dependence on the charm-quark mass $m_c$,
      and consequently the partonic phase-space integration for the calculation
      of various moments of the differential rate has to be modified
      accordingly, in order to avoid double counting. 

\end{enumerate}
In the following we shall discuss the different cases in turn.

\subsection{ $\mathbf {m_b \sim m_c \gg \Lambda_{\rm QCD}}$ } 

As explained above, when integrating out both, the hard $b$\/-quark fluctuations
and the charm quark, at a common scale $\mu \sim m_b$, 
we are left with operators built from soft fields, 
only.  
Thus the only matrix elements appearing at order $1/m_b^3$ are the Darwin term $\rho_D^3$ and 
the spin-orbit term $\rho_{LS}^3$ defined by (we use the convention of \cite{Dassinger:2006md}, but omit the hat over $\hat\rho_D^3,\hat \rho_{LS}^3$)
\begin{align}
2 M_B \, \rho_D^3 &= \langle \bar B(p) | \bar{b}_v \, (iD_\mu) (ivD) (iD^\mu) \, b_v  | \bar B(p) \rangle
\,, \cr  
2 M_B \, \rho_{LS}^3 &= \langle \bar B(p) | \bar{b}_v \, (iD_\mu) (ivD) (iD_\nu) (-i \sigma^{\mu \nu}) \, b_v  | \bar B(p) \rangle \,.
\end{align}
In the charged-lepton energy spectrum one obtains (among others) a contribution of the form 
\begin{equation}
\frac{d\Gamma}{dy} \Big|_{\rho_D^3} = 
   \frac{G_F^2 m_b^5}{192 \pi^3} |V_{cb}|^2 \, \frac{\rho_D^3}{m_b^3} 
   \bigg\lbrace - \frac{8 \, \theta(1-y-\rho)}{1-y} + \ldots \bigg\rbrace \,.
\label{case1dGamma}
\end{equation}
For later use, we have only quoted the most singular term in the limit
$y = 2 E_\ell /m_b \to 1$, and $\rho = m_c^2 / m_b^2 \to 0$
(the full expressions are provided in the appendix).
Upon integration it yields a logarithmically enhanced
contribution to the total rate
\begin{equation}
\Gamma \Big|_{\rho_D^3}=  \frac{G_F^2 m_b^5}{192 \pi^3} \, |V_{cb}|^2 \,
 \frac{\rho_D^3}{m_b^3} \left\{ 8 \, \ln\rho + \ldots \right\} \,,
\label{Gamma3}
\end{equation}
where the ellipses denote the contributions from the
sub-leading terms in (\ref{case1dGamma}) which are of order $\rho \, \ln\rho$.
Similarly, we identify the leading terms in the moments (see appendix)
\begin{align}
 \langle (y-y_0)^n \rangle \Big|_{\rho_D^3}
 &= \frac{G_F^2 m_b^5}{192 \pi^3} \, |V_{cb}|^2 \,
 \frac{\rho_D^3}{m_b^3} \left\{  8 \, (1-y_0)^n \, \ln \rho + \ldots \right\} \label{ymom3log}\,.
\end{align}
Note that for $m_b \sim m_c$ the logarithm is actually of order one and represents 
a regular contribution to the matching coefficient (and therefore the remaining
terms in curly brackets enter on the same level).
Also, the phase space boundary for $y$ is $y < 1- \rho$ which is 
away from $y = 1$ by an amount of order one. 

A similar logarithmically enhanced term also appears in the partonic
rate,
\begin{equation}
\Gamma \Big|_{\rm partonic}=  \frac{G_F^2 m_b^5}{192 \pi^3} \, |V_{cb}|^2 \,
  \left\{ 1 - 12 \, \rho^2 \, \ln\rho + \ldots \right\} \,,
\label{Gamma0}
\end{equation}
and in the related moment,
\begin{equation}
\langle 1-y \rangle \Big|_{\rm partonic}=  \frac{G_F^2 m_b^5}{192 \pi^3} \, |V_{cb}|^2 \,
  \left\{ 6 \, \rho^2 \, \ln\rho + \ldots \right\} \,.
\label{ymom0}
\end{equation}
In contrast to the Darwin-term contribution, the logarithmic term
vanishes in the limit $\rho \to 0$. Nevertheless, as has been shown in
\cite{Bauer:1996ma}, such ``phase-space logs'' can be resummed into
short-distance coefficients, as we are going to discuss in the following.


\subsection{ $\mathbf {m_b \gg m_c \gg \Lambda_{\rm QCD}}$} 
When we integrate out the $b$ quark first at a scale $\mu_h \sim m_b$ 
and still keep the charm quark dynamical,
we have to take into account operators with explicit charm quarks until 
those are finally integrated out at the semi-hard scale $\mu_{sh} \sim m_c$. 
In addition to the dimension-5 and dimension-6 operators, defining $\mu_\pi^2,\mu_G^2$
and $\rho_D^3,\rho_{LS}^3$,
one thus finds (at tree level) matrix elements of 
the local ``intrinsic-charm'' operators
\begin{align}
2M_B \, W_{\mu \nu}^{IC}& = 
( 2\pi)^4 \, \delta^4 (q-m_b v) \, \langle \bar B(p) |
(\bar{b}_v  \, \gamma_\nu P_L
\,  c)
\, ( \bar{c}  \, \gamma_\mu  P_L  \, b_v ) | \bar B(p) \rangle 
\cr
& + ( 2\pi)^4 \left(\frac{\partial}{\partial q_\alpha} 
 \, \delta^4 (q-m_b v) \right) \ 
\langle \bar B(p) |(i\partial_\alpha \, \bar{b}_v  \, \gamma_\nu P_L
\,  c) \, (\bar{c}  \, \gamma_\mu  P_L  \, b_v)  | \bar B(p) \rangle 
\cr 
& + \ldots
\,,
\label{WIC}
\end{align} 
which can be interpreted as the probability to find semi-hard (i.e.\ off-shell) 
charm quarks inside the heavy $\bar B$\/-meson.
 
Notice, that the power-counting for the semi-hard charm fields 
$[c]=(m_c)^{3/2}$ is now 
different from the ones for soft HQET fields $[b_v] = \Lambda^{3/2}$, 
and therefore it may be convenient 
to use a notation as in \cite{Boos:2005by}, where
the ``intrinsic-charm'' operators in the first line of (\ref{WIC})
are suppressed by $\lambda^3 \equiv (m_c/m_b)^{3}$,
the ones in the second line by $\lambda^4$,
the kinetic and chromomagnetic operators by $\lambda^4 \equiv (\Lambda/m_b)^2 $, and the Darwin and
spin-orbit term by $\lambda^6$. Due to chiral symmetry, 
only the $\lambda^4$ ``intrinsic-charm'' operators contribute to the
partonic rate for $b \to c\ell \nu$, related to the
$\rho^2 \, \ln\rho$ term in (\ref{Gamma0}).
Additional soft gluon couplings to semi-hard charm quarks
are further suppressed, and this will give 
rise to the $\lambda^6$ suppressed terms $\rho_D^3 \ln \rho$ 
in (\ref{Gamma3}), descending from the $\lambda^3$ ``intrinsic-charm'' operators.

Let us consider first
the matrix elements of the operator in the first line of (\ref{WIC}).
They may be decomposed 
in terms of two hadronic parameters, $T_1(\mu)$ and $T_2(\mu)$, 
\begin{equation} \label{IC} 
(4\pi)^2 \, \langle \bar B(p) |\bar{b}_v \, \gamma_\nu P_L \, c
\ \bar{c} \, \gamma_\mu  P_L \, b_v  | \bar B(p) \rangle  
=  2 M_B \left( T_1(\mu) \, g_{\mu \nu} + T_2(\mu) \,v_\mu v_\nu \right) \,.
\end{equation}

The contribution to the rate of the matrix element of the local 
``intrinsic-charm'' operators is concentrated at small hadronic mass $m_X$ and 
in the endpoint of the lepton energy spectrum. Performing the tree-level
matching at $\mu=m_b$, we have 
\begin{equation} \label{spec} 
\frac{d^2 \Gamma^{IC} }{dm_X^2 \, dy} = \delta(m_X^2) \, \delta ( 1-y ) \, \Gamma^{IC} \quad 
\mbox{ and  } \quad
\frac{d  \Gamma^{IC} }{dy} = \delta ( 1-y ) \, \Gamma^{IC} \, ,
\end{equation}
with
\begin{equation}
  \Gamma^{IC} =  -   \frac{G_F^2 m_b^5}{24\pi^3} \, |V_{cb}|^2 \ 
      \frac{3 \,  T_1(m_b)}{m_b^3} \,. 
\label{case2dGammaIC}
\end{equation}

On the other hand, the calculation of the matching coefficients for 
the contribution of $\rho_D^3$  and $\rho_{LS}^3$ to the total rate
now has to be performed in the limit $m_c \ll m_b$. 
Notice, that the naive limit $\rho \to 0$ in (\ref{case1dGamma}) would
give ill-defined expressions.
In particular, the integral over
\[
  dy \, \frac{\theta(1-y)}{1-y}
\]
would be infrared divergent in the lepton-energy endpoint.
As we will see, the new IR divergence in the phase-space integration, appearing
in the limit $\rho \to 0$, is related to the UV renormalization 
of the ``intrinsic-charm'' operators (\ref{IC}).
Defining the hadronic parameters $T_{1,2}(\mu)$ in the $\overline{\rm MS}$ scheme,
we also have to perform the phase-space integral in $D=4-2\epsilon$ dimensions.
As a result, the contribution of the Darwin term to the total rate 
is regularized by plus-distributions,
\begin{equation}
 \frac{\theta(1-y)}{1-y} \to 
\left[ \frac{\theta(1-y)}{1-y} \right]_+ - \delta(1-y) \, 
 \ln \left(\frac{\mu^2}{m_b^2} \right) \,,
\end{equation}
which exactly subtracts the effects of semi-hard charm quarks,
that would otherwise be double-counted when adding (\ref{case2dGammaIC})
to the decay rate.
 
The final expression for the combined contributions of the Darwin term and the 
``intrinsic-charm'' operators to the lepton-energy 
spectrum at order $1/m_b^3$ can be written
as
\begin{align}
\frac{d \Gamma^{(3)}}{dy}
\Big|_{\rho_D^3 + \rm IC} & = 
   \frac{G_F^2 m_b^5}{24 \pi^3} \, |V_{cb}|^2 
   \bigg\lbrace \frac{C_{\rho_D}(y,\mu) \, \rho_D^3(\mu)}{m_b^3}  
    +   \frac{ C_{T_1}(y,\mu) \, T_1(\mu)}{m_b^3} 
\bigg\rbrace \,,
\label{case2dGamma}
\end{align}
which should be used for $ m_c \leq \mu \leq m_b$.
The matching conditions for the short-distance
coefficient functions --
including the limit $\rho \to 0$ for the
sub-leading terms in (\ref{case1dGamma}) as given in the appendix
 -- 
are given by
\begin{align}
C_{\rho_D}(y,m_b) & =
 -  \left[ \frac{y^2 (9-5y+2y^2) \, \theta(1-y)}{6(1-y)}\right]_+  + \frac{17}{12} \, \delta(y-1) 
\cr & \quad 
 + \frac{5}{24} \, \delta'(y-1) 
 - \frac{1}{72} \, \delta''(y-1) + {\cal O}(\alpha_s) \,,
\nonumber \\[0.2em]
C_{T_1}(y,m_b) & =  - 3 \, \delta(y-1) + {\cal O}(\alpha_s) \,,
\cr 
C_{T_2}(y,m_b) & = {\cal O}(\alpha_s) \,.
\label{match}
\end{align}

\begin{figure}[t!pt]
 \centering
  \includegraphics[width=0.8\textwidth]{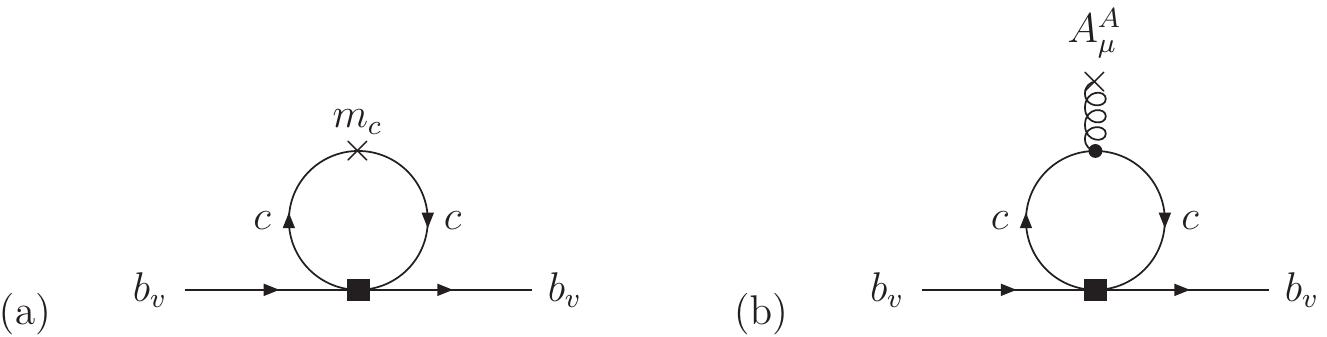}
\parbox{0.8\textwidth}{
\caption{\small Leading diagrams determining the mixing of four-quark
         into two-quark operators. \label{figb}}}
\end{figure}

In appendix \ref{mixing}, we derive the leading terms
in the anomalous dimension matrix that describe the 
mixing of the ``intrinsic-charm'' operators 
$\{  T_1(\mu), \, T_2 (\mu) \}$
into the Darwin term 
$\rho_D(\mu)$, see also Fig.~\ref{figb}(b),
\begin{align}
 \frac{d}{d\ln\mu} \begin{pmatrix} \rho_D(\mu) \,\, 
\\  T_1(\mu) \\   T_2(\mu) \end{pmatrix}
= - \left\{ \begin{pmatrix}
       0 & 0 & 0 \\ 
       -2/3 & 0 & 0 \\ 
       \phantom{-}4/3 & 0 & 0                                           
       \end{pmatrix}
 + {\cal O}(\alpha_s) \right\} \begin{pmatrix} \rho_D(\mu) \,\, 
\\ T_1(\mu) \\ T_2(\mu) \end{pmatrix} \,.
\end{align}
Neglecting the ${\cal O}(\alpha_s)$ contributions to the anomalous dimension
matrix, we only determine the leading-logarithmic terms,\footnote{Strictly speaking, these are $\rm N^{-1}LL$. To resum the
leading-logarithms of the order 
$\alpha_s^n \times \ln^n\rho$, we would need the ${\cal O}(\alpha_s)$
mixing of the Darwin term into itself and the complete set of
``intrinsic-charm'' operators into themselves, which goes beyond the
scope of this work (see however \cite{Bauer:1996ma} for a complete
leading-log analysis for the dimension-7 contributions to the
total rate).} which are generated 
by the renormalization-group equation for the short-distance coefficients
\begin{align}
C_{T_i}(y,\mu) & \simeq C_{T_i}(y,m_b) \,,
\nonumber \\[0.1em]  
C_{\rho_D}(y,\mu) & \simeq 
 C_{\rho_D}(y,m_b) - \frac{1}{3} \, \ln \frac{\mu^2}{m_b^2}
\, \left( C_{T_1}(y,m_b) - 2 C_{T_2}(y,m_b) \right)
\,. 
\label{Cresum}
\end{align}
Now, integrating out the semi-hard charm quarks at $\mu_{\rm sh} = m_c$,
is equivalent to setting 
\begin{equation}
  T_i(\mu \leq m_c) = 0 \,.
\label{LOmatch}
\end{equation}
In this case, the expression for the lepton-energy spectrum (\ref{case2dGamma})
simplifies to
\begin{align}
\frac{d \Gamma^{(3)}}{dy}
\Big|_{\rho_D^3 + \rm IC} & = 
   \frac{G_F^2 m_b^5}{24 \pi^3} \, |V_{cb}|^2 \,
   \frac{C_{\rho_D}(y,m_c) \, \rho_D^3(m_c)}{m_b^3}  
\,,
\label{case2dGamma2}
\end{align}
and the information on ``intrinsic charm'', i.e.\ the non-analytic
dependence on the charm-quark mass, has been completely absorbed into
the short-distance function $C_{\rho_D}(y,m_c)$.
This can be made explicit by inserting the leading-order matching
conditions (\ref{match}) for $C_{T_i}(y,m_b)$, which results in
\begin{align}
 	C_{\rho_D}(y,m_c) &
 \simeq C_{\rho_D}(y,m_b) + \ln \frac{m_c^2}{m_b^2} \ \delta(y-1)
\,.
\end{align}
In this way (\ref{case2dGamma2}) reproduces the logarithmic
term in the lepton-energy moments in (\ref{ymom3log}) as well
as the finite terms 
(given by the limit $\rho \to 0$ of Eq.~(\ref{ymom3}) in the
Appendix).

Similar considerations can be made for the $\rho^2 \, \ln\rho$ term
in the partonic rate. 
We decompose the matrix elements of the operators in the
second line of (\ref{WIC}) as
\begin{align} \label{IC2} 
& 
(4\pi)^2 \, \langle \bar B(p) | ( i\partial_\alpha \, \bar{b}_v \, \gamma_\nu P_L \,  c)
\, (\bar{c} \, \gamma_\mu  P_L \, b_v)  | \bar B(p) \rangle  
\cr 
= & \, 2 M_B \left( 
T_3(\mu) \, g_{\mu \nu} \, v_\alpha + 
T_4(\mu) \, g_{\mu \alpha} \, v_\nu  +
T_5(\mu) \, g_{\nu \alpha} \, v_\mu  
 +
T_6(\mu) \, v_\mu v_\nu v_\alpha -
T_7(\mu) \, i \epsilon_{\mu\nu\alpha\beta} v^\beta
\right) \,.
\cr 
\end{align}
(Notice that in unpolarized observables, only the sum $T_4(\mu)+T_5(\mu)$ appears.)
Generalizing the results for the total rate
in \cite{Bauer:1996ma} to the lepton-energy spectrum, and concentrating
again on the leading logarithmic terms, we find 
\begin{align}
\frac{d \Gamma}{dy}
\Big|_{\rm partonic + IC} & = 
   \frac{G_F^2 m_b^5}{192 \pi^3} \, |V_{cb}|^2 
   \bigg\lbrace 
C_0(y,\mu) + \rho \, C_1(y,\mu) + \rho^2 \, C_2(y,\mu)
\cr & \qquad \qquad \qquad 
    +  \frac{ \sum_{i=3}^7 \, C_{T_i}(y,\mu) \, T_i(\mu)}{m_b^4} 
\bigg\rbrace \,,
\label{case2dGamma0}
\end{align}
with
\begin{align}
C_0(y,m_b) & = \left( 6 y^2 - 4 y^3 \right) \theta(1-y)+ {\cal O}(\alpha_s)\,, 
\cr 
C_1(y,m_b) & = - 6 y^2 \, \theta(1-y) - 6 \, \delta(y-1)+ {\cal O}(\alpha_s) \,,
\cr 
C_2(y,m_b) &=
\left[
\frac{ 12 \, \theta(1-y)}{1-y}
\right]_{+}
- 
\left[
 \frac{6 \, \theta(1-y)}{(1-y)^2}
\right]_{++}
\cr & \qquad
- 6 \, \theta(1-y)
+ 6 \, \delta(y-1)
+ 3 \, \delta'(y-1) + {\cal O}(\alpha_s) 
 \,,
\end{align}
and
\begin{align}
C_{T_3}(y,m_b) &=  - 24 \,\delta'(y-1) + 48 \,\delta(y-1) + {\cal O}(\alpha_s) \,, 
\cr 
C_{T_4,T_5}(y,m_b) &= - 24 \,\delta(y-1) + {\cal O}(\alpha_s) \,, 
\cr 
C_{T_6}(y,m_b)&= {\cal O}(\alpha_s) \,,
\cr 
C_{T_7}(y,m_b)&=   24 \,\delta'(y-1) + {\cal O}(\alpha_s) \,.
\end{align}
Again, the ``intrinsic-charm'' operators $T_{3-7}$ mix into the
2-particle operator $m_c^4 \, \bar b \, v_\mu \gamma^\mu \, b$
(see appendix), and consequently, the coefficient $C_2(y,\mu)$
evolves as
\begin{align}
C_2(y,m_c) & \simeq 
  C_2(y,m_b) 
 - \frac{1}{8} \,
   \ln\frac{\mu^2}{m_b^2} 
               \left(C_{T_3}(y,m_b) -  C_{T_4}(y,m_b) - C_{T_5}(y,m_b) - C_{T_7}(y,m_b) \right) 
 \,.
\label{C2}
\end{align}
Inserting the leading-order matching conditions, one has
\begin{align}
 - \frac18 \left( C_{T_3}(y,m_b) -  C_{T_4}(y,m_b) - C_{T_5}(y,m_b) - C_{T_7}(y,m_b)
 \right)  &= 6 \,\delta'(y-1) - 12 \, \delta(y-1) \,,
\end{align}
and one reproduces the logarithmic terms
$-12 \rho^2 \ln\rho $ in $\Gamma_{\rm part}$ and $6\rho^2 \ln \rho$ 
in $\langle 1-y\rangle_{\rm part}$, respectively, see
(\ref{Gamma0},\ref{ymom0}).

\subsection{ $\mathbf {m_b \gg m_c \sim \Lambda_{\rm QCD}}$} 

If we consider the dynamics at the charm-quark mass scale to be
in the non-perturbative regime,
we cannot exploit the condition (\ref{LOmatch}) and are
left with the general formula for the leptonic-energy spectrum
in (\ref{case2dGamma}), which should be evaluated at a scale
$\mu_0$ that satisfies $m_c \ll \mu_0 \leq m_b$. Moreover, we
have to take seriously the new power counting which implies that
terms of order $\rho^2$ now count as $(\Lambda/m_b)^4$ and should
be neglected to the order that we are considering, we are thus left 
with
\begin{align}
 \frac{d \Gamma}{dy}
\Big|_{\rm partonic} & = 
   \frac{G_F^2 m_b^5}{192 \pi^3} \, |V_{cb}|^2 
   \bigg\lbrace 
C_0(y,\mu_0) + \rho \, C_1(y,\mu_0) + {\cal O}(\rho^2) \bigg\rbrace \,,
\label{dGamma3a}
\\[0.1em] 
\frac{d \Gamma^{(3)}}{dy}
\Big|_{\rho_D^3 + \rm IC} & = 
   \frac{G_F^2 m_b^5}{24 \pi^3} \, |V_{cb}|^2 
   \bigg\lbrace \frac{C_{\rho_D}(y,\mu_0) \, \rho_D^3(\mu_0)}{m_b^3}  
    +   \frac{ C_{T_1}(y,\mu_0) \, T_1(\mu_0)}{m_b^3} 
\bigg\rbrace \,,
\label{dGamma3b}
\end{align}
together with the contributions to the lepton-energy spectrum
from $\mu_\pi^2,\mu_G^2$ and
$\rho_{\rm LS}^3$ (see e.g.\ \cite{Dassinger:2006md}), 
where the limit $\rho \to 0$ to the considered order ($1/m_b^3$) is trivial.

In that order, the genuinely intrinsic-charm contribution comes
together with the Darwin term, only.
In particular, to leading logarithmic accuracy (\ref{Cresum}),
the contributions to the total rate, and the moments $\langle y \rangle$
and $\langle y^2 \rangle$ can be obtained as
\begin{align}
\Gamma^{(3)}
\Big|_{\rho_D^3 + \rm IC} & = 
\frac{G_F^2 m_b^5}{24 \pi^3} \, |V_{cb}|^2 \,
   \Bigg\{ X(\mu_0)
   + \frac{\rho_D^3(\mu_0)}{m_b^3}  \left[ \frac{17}{12}\right] 
\Bigg\}
\,, 
\label{i}
\\
\langle y \rangle \, \Big|_{\rho_D^3 + \rm IC} & = 
\frac{G_F^2 m_b^5}{24 \pi^3} \, |V_{cb}|^2 \,
   \Bigg\{ X(\mu_0)
   + \frac{\rho_D^3(\mu_0)}{m_b^3}  \left[ \frac{47}{30}\right] 
\Bigg\}
\label{ii}
\,, \\
\langle y^2 \rangle \, \Big|_{\rho_D^3 + \rm IC} & =  
\frac{G_F^2 m_b^5}{24 \pi^3} \, |V_{cb}|^2 \,
   \Bigg\{ X(\mu_0)
   + \frac{\rho_D^3(\mu_0)}{m_b^3}  \left[ \frac{287}{180} \right]
\Bigg\}
\label{iii}
\,,
\end{align}
where we defined the parameter combination
\begin{equation}
X(\mu_0) =  
- \frac{3  \, T_1(\mu_0)}{m_b^3}
   + \ln \frac{\mu_0^2}{m_b^2} \, \frac{\rho_D^3(\mu_0)}{m_b^3} 
\,.
\end{equation}
Considering a sizeable value for $T_1(\mu_0)$ at small hadronic scales
(in contrast to the perturbative situation considered in the previous
subsection), and taking into account that the $\rho_D^3$ contribution in $X(\mu_0)$
is formally enhanced by $\ln \mu_0^2/m_b^2$, 
we may ignore the (small) differences between the individual moments
induced by the numbers in square brackets in (\ref{i},\ref{ii},\ref{iii}), to first
approximation.
Therefore, even in this genuine intrinsic-charm scenario, 
the inclusion of a large non-perturbative intrinsic-charm
effect, basically amounts to treating the Darwin term
$\rho_D^3$ for the effective parameter $X$.
In any case, one may consider the limit $m_c \sim \Lambda_{\rm QCD}$ rather academic,
and would prefer the scenario with semi-hard charm-quarks as in the previous
subsection for the discussion of  ``intrinsic-charm'' effects in inclusive
semi-leptonic $B$ decays.

We should also mention that (\ref{dGamma3a},\ref{dGamma3b})  provide the
appropriate formulas for the massless limit, relevant to $b \to u \ell \nu$ 
decays, after appropriate changes $V_{cb} \to V_{ub}$ and re-interpretation
of the intrinsic-charm operators as so-called weak  annihilation operators 
\cite{Bigi:1993bh,Gambino:2008fj}.
Notice that the (local) annihilation operators enter at order $1/m_b^3$
in the standard OPE, whereas their non-local counterparts, necessary to
describe the shape-function region, already enter at (relative) 
order $\Lambda/m_b$ 
\cite{Lee:2004ja,Bosch:2004cb,Beneke:2004in}.

\section{Conclusion} 
We have shown how the ``intrinsic-charm'' contribution in semi-leptonic $B$\/-meson decays  is related to the renormalization of 
sub-leading operators (like $m_c^4 \, \bar b_v b_v$
and the Darwin term) appearing in the operator product expansion for the lepton-energy spectrum and the total rate. We have distinguished three different cases which correspond
to different power counting for the charm-quark mass. In the first case, one
assumes $m_b \sim m_c$, i.e.\ the charm-quark is already integrated out
at the hard scale, set by the large $b$\/-quark mass in the OPE.
Consequently, all dependence on the charm-quark dynamics is already encoded in
the matching conditions for the hard coefficient functions, and no ``intrinsic-charm''
operators should be introduced below the hard scale. The only remnant of 
``intrinsic charm'' is the non-analytic dependence of the coefficient functions
on the ratio $\rho =m_c^2/m_b^2$.

Another viable scenario treats the charm quark mass as intermediate between
the hard and the soft scale in the OPE, $m_b \gg m_c \gg \Lambda_{\rm QCD}$.
In that case, four-quark operators including soft $b$\/-quark fields and
semi-hard charm quarks have to be included in the OPE. At the same time,
in order to avoid double counting, the semi-hard region has to be subtracted
from phase-space integrals by a suitable regularization of the decay spectra
in the limit $m_c \ll m_b$. We have shown by explicit calculation how the
mixing between the ``intrinsic-charm'' operators and the Darwin term  
generates the logarithmically enhanced terms entering the OPE at 
order $1/m_b^3$. Similarly, extending the results of \cite{Bauer:1996ma},
we could reproduce terms of order $\rho^2 \ln \rho$ in the partonic rate.
After integrating out the charm quark at the semi-hard scale, the moments of
the lepton-energy spectrum can be entirely described in terms of the
standard hadronic input parameters, whereas -- again -- the complete charm-quark
dependence enters via (eventually renormalization-group improved) short-distance
coefficients, multiplying, for instance, the Darwin term.

A somewhat more exotic approach would treat the charm quark as light,
i.e.\ of order $\Lambda_{\rm QCD}$. Only in this case genuine intrinsic-charm
(i.e.\ non-perturbative) effects have to be taken into account. 
Still, we have found that
on the level of a few lepton-energy moments, the experimental data 
basically constrain a particular combination of the intrinsic-charm
contribution and the Darwin term, such that to order $1/m_b^3$ the
number of independent hadronic parameters effectively remains the same.

The main conclusion to be drawn is that, as long as the strong dynamics
at the charm-quark scale can be treated perturbatively,
``intrinsic-charm'' effects do not induce an additional source of hadronic 
uncertainties  at the level of $1/m_b^3$ power corrections, apart from
the usually considered Darwin and spin-orbit terms. The same will be true
for higher orders in the $1/m_b$ expansion as classified
in \cite{Dassinger:2006md}.
The issue of whether to resum logarithms $\ln (m_c^2 / m_b^2)$ by introducing the above
2-step matching procedure, or sticking to the standard 1-step matching
has to be decided by considering radiative corrections to the $1/m_b^3$
expressions which is beyond the scope of this work (see also \cite{Pak:2008qt}).

\subsection*{Acknowledgements}
This work is partially supported by
the German Research Foundation (DFG) under Contract 
No.~MA1187/10-1, and by the German Ministry of Research (BMBF,
Contract No.~05HT6PSA).

\appendix

\section{Lepton-Energy Spectrum and Moments}

\subsection{Partonic rate}

The complete expression for the partonic contribution to
the lepton-energy spectrum  with  $m_b \sim m_c \geq \mu$ 
is given by \cite{Dassinger:2006md}
\begin{align}
 \frac{d\Gamma}{dy} \Big|_{\rm partonic} = &
   \frac{G_F^2 m_b^5}{192 \pi^3} |V_{cb}|^2 
   \bigg\lbrace 
\Big(
-\frac{4 \rho ^3}{(y-1)^3}
-\frac{6 \left(\rho ^3+\rho ^2\right)}{(y-1)^2}
-\frac{12 \rho ^2}{y-1}
\cr 
& \qquad
-4 y^3
-6 (\rho -1) y^2
+2 (\rho -3) \rho^2
\Big) \theta(1-y-\rho)
\bigg\rbrace \,.
\label{dGammapart}
\end{align}
From this one can obtain closed expressions for the $(1-y)^n$ moments,
\begin{align}
\langle (1-y)^n \rangle\Big|_{\rm partonic} = &
   \frac{G_F^2 m_b^5}{192 \pi^3} |V_{cb}|^2 
   \bigg\lbrace
-\frac{12 \left(\rho ^n-1\right) \rho ^2}{n}-\frac{4 \left(\rho ^n-\rho
   ^2\right) \rho }{n-2}-\frac{12 \left(\rho ^{n+2}-1\right) \rho
   }{n+2}
\cr & \qquad   
+\frac{6 \left(\rho ^2+\rho \right) \left(\rho ^n-\rho
   \right)}{n-1}-\frac{2 \left(\rho ^3-3 \rho ^2-3 \rho +1\right)
   \left(\rho ^{n+1}-1\right)}{n+1}
\cr & \qquad   
+\frac{6 (\rho +1) \left(\rho
   ^{n+3}-1\right)}{n+3}-\frac{4 \left(\rho ^{n+4}-1\right)}{n+4}
\bigg\rbrace \,.
\label{ymompart}
\end{align}
Arbitrary moments can be derived via
\begin{align}
\langle (y-y_0)^n \rangle & = \sum_{k=0}^n 
\begin{pmatrix} n \, \\ k\, \end{pmatrix}
 (1-y_0)^{n-k} \,(-1)^k \, \langle (1-y)^{k} \rangle \,.
\end{align}

Expanding (\ref{ymompart}) in the small parameter $\rho$, the
logarithmically enhanced terms at order $\rho^2$ appear only in the
total rate and the first moment 
\begin{align}
\langle (1-y)^n \rangle\Big|_{\rm partonic} = &
 \frac{G_F^2 m_b^5}{192 \pi^3} |V_{cb}|^2 
\left( 12 \, \delta^{n0} - 6 \, \delta^{n1} \right) \rho^2 \ln \rho 
+ \mbox{\footnotesize analytic/higher-order terms in $\rho$.}
\end{align}

\subsection{Darwin-term contribution}

The full contribution related to the Darwin term in the
lepton-energy spectrum for the case $m_b \sim m_c \geq \mu$ 
is given by \cite{Dassinger:2006md}
\begin{align}
 \frac{d\Gamma^{(3)}}{dy} \Big|_{\rho_D^3} = &
   \frac{G_F^2 m_b^5}{192 \pi^3} |V_{cb}|^2 \, \frac{\rho_D^3}{m_b^3} 
   \bigg\lbrace \nonumber \\
& \qquad \left( \frac{40 \rho ^3}{3 (y-1)^6}+\frac{8 \rho ^2 (3 \rho
   +1)}{(y-1)^5}+\frac{6 \rho ^2 (3 \rho +1)}{(y-1)^4}+\frac{16 \rho 
   \left(2 \rho ^2-\rho -1\right)}{3 (y-1)^3}\right.
\cr 
& \qquad \left. -\frac{28 \rho }{3
   (y-1)^2}+\frac{8}{y-1}+\frac{2}{3} \left(5 \rho ^3-5 \rho ^2+10 \rho
   +22\right) \right. 
\cr 
& \qquad \left. +\frac{8}{3} (\rho +3) (y-1)+4 (y-1)^2+\frac{8}{3}
   (y-1)^3 
\right) \theta(1-y-\rho)
\cr
& \qquad - \left( \frac{2 (\rho -1)^4 (\rho +1)^2}{3 \rho ^2}\right) \delta(1-y-\rho)
\bigg\rbrace
\,.
\label{dGamma3}
\end{align}
From this one can obtain closed expressions for the moments,
\begin{align}
\langle (1-y)^n \rangle\Big|_{\rho_D^3} = &
   \frac{G_F^2 m_b^5}{192 \pi^3} |V_{cb}|^2 \, \frac{\rho_D^3}{m_b^3} 
   \bigg\lbrace
\frac{8 \left(\rho^n-1\right)}{n}
-\frac{2}{3} (\rho -1)^4 (\rho +1)^2 \rho ^{n-2}+\frac{28 \left(\rho ^n-\rho \right)}{3 (n-1)}
\cr & \qquad   
-\frac{2
   \left(5 \rho ^3-5 \rho ^2+10 \rho +22\right) \left(\rho
   ^{n+1}-1\right)}{3 (n+1)}+\frac{8 (\rho +3) \left(\rho
   ^{n+2}-1\right)}{3 (n+2)}
\cr & \qquad   
-\frac{4 \left(\rho
   ^{n+3}-1\right)}{n+3}+\frac{8 \left(\rho ^{n+4}-1\right)}{3
   (n+4)}
-\frac{16 \left(2 \rho ^2-\rho -1\right) \left(\rho ^2-\rho
   ^n\right)}{3 (n-2) \rho }
\cr & \qquad   
+\frac{6 (3 \rho +1) \left(\rho ^3-\rho
   ^n\right)}{(n-3) \rho }-\frac{8 (3 \rho +1) \left(\rho ^4-\rho
   ^n\right)}{(n-4) \rho ^2}+\frac{40 \left(\rho ^5-\rho ^n\right)}{3
   (n-5) \rho ^2}
\bigg\rbrace \,.
\label{ymom3}
\end{align}
Taking the limit $\rho \to 0$ in (\ref{ymom3}), the
logarithmically enhanced terms  appear only  in the
total rate 
\begin{align}
\langle (1-y)^n \rangle\Big|_{\rho_D^3} = &
\frac{G_F^2 m_b^5}{24 \pi^3} |V_{cb}|^2 \, \frac{\rho_D^3}{m_b^3}  
\left( \delta^{n0} \,  \ln \rho 
+ {\cal O}(\rho \ln\rho) \right) \,.
\end{align}

\section{Operator Mixing}

\label{op_basis}

\subsection{Dimension-6}

In the following we briefly sketch the derivation of the
elements of the anomalous-dimension matrix that govern
the mixing of the four-quark (``intrinsic-charm'') operators
into the Darwin term. For simplicity, we do not construct
the complete set of independent operators that would be needed
to describe the full one-loop anomalous-dimension matrix, but
rather focus on the effect of the charm-loop diagram 
in Fig.~\ref{figb}(b). For this purpose it is sufficient to
consider the two operator structures which enter the hadronic
tensor at tree-level (\ref{WIC}):
\begin{align}
 	2M_B T_1(\mu)  &=  
\frac{(4\pi)^2}{3} \Big( \langle \bar B(p) |\bar{b}_v  \, \gamma_\mu P_L  c \ 
          \bar{c} \, \gamma^\mu P_L \,  b_v  | \bar B(p) \rangle  - 
 \langle \bar B(p) |\bar{b}_v \, \slashed{v} P_L \, c \
               \bar{c} \, \slashed{v}  P_L \,  b_v  | \bar B(p) \rangle \Big)\,,
 \cr 
 2M_B T_2(\mu) &=  \frac{(4\pi)^2}{3} \Big( 4\, \langle \bar B(p) |\bar{b}_v  \, \slashed{v} P_L \, c
      \ \bar{c} \, \slashed {v} P_L \,  b_v  | \bar B(p) \rangle  - 
\langle \bar B(p) |\bar{b}_v  \, \gamma_\mu P_L \,  c 
      \ \bar{c} \, \gamma^\mu  P_L \,  b_v  | \bar B(p) \rangle \Big)\,.
\end{align}
Together with the Darwin term they are used to define a simplified operator
basis
\begin{align}
 {\cal O}_{\rho_D} &= \bar{b}_v \, (iD_\mu) (ivD) (iD^\mu) \, b_v \,, 
\cr 
 {\cal O}_{T_1} &= (4\pi)^2 \, \mu^{2\epsilon} \,
\frac13 \left( \bar{b}_v \, \gamma_\mu P_L \, c \ \bar{c}\, \gamma^\mu P_L \, b_v -
\bar{b}_v \, \slashed{v} P_L \,  c \ \bar{c} \, \slashed{v} P_L \,  b_v \right)
\,, 
\cr 
 {\cal O}_{T_2} &= (4\pi)^2 \, \mu^{2\epsilon} \,
\frac13 \left(4 \,
\bar{b}_v \, \slashed{v} P_L \,  c \ \bar{c} \, \slashed{v} P_L \,  b_v 
 - \, \bar{b}_v \, \gamma_\mu P_L \, c \ \bar{c}\, \gamma^\mu P_L \, b_v \right)
  \,.
\end{align}
Notice that for convenience, we have extracted a factor $(4\pi)^2 \, \mu^{2\epsilon}$, in order to have a simple, dimensionless 
anomalous-dimension matrix.\footnote{With this convention, the anomalous-dimension matrix is of order $(\alpha_s)^0$.
In order to have it in the standard form, one would have to extract
a factor $g_s^2 = 4 \pi \alpha_s \, \mu^{2\epsilon}$, instead.}


\label{mixing}

Calculating the one-loop matrix elements of the operators ${\cal O}_{T_{1,2}}$
for the partonic transition $b \to b$ in the presence of a soft
background field $A_\mu(k)$, see Fig.~\ref{figb}(b), and comparing with the tree-level
matrix element of the Darwin-term operator, we obtain the following results
in $D=4-2\epsilon$ dimensions,
\begin{align}
\langle b| {\cal O}_{T_1}|b \rangle^{(0)} 
     &= +\frac{1}{3}  \left(\frac{1}{\epsilon}+\ln \frac{\mu^2}{m_c^2}\right) \langle b|{\cal O}_{\rho_D} |b \rangle_{\rm tree}
\,,
 \cr
\langle  b|{\cal O}_{T_2}|b  \rangle^{(0)}
     &= - \frac{2}{3}  \left(\frac{1}{\epsilon}+\ln \frac{\mu^2}{m_c^2}\right)\langle b|{\cal O}_{\rho_D} |b \rangle_{\rm tree} 
\,,
\label{div}
\end{align}
where the one-gluon matrix element of the Darwin-term operator 
on parton level is given by
\begin{align}
 \langle b|{\cal O}_{\rho_D} |b \rangle_{\rm tree} & = 
\frac12 \, \langle b| \bar b_v \left[ i D_\mu, \left[(i v \cdot D), iD^\mu\right]\right] b_v|b \rangle_{\rm tree} + {\cal O}(1/m_b)
\cr 
& = \frac{g}{2} \left( (v\cdot k)(k \cdot A) - k^2 \, (v \cdot A) \right) \bar u_b \, u_b
  + \ldots
\end{align}
From (\ref{div}) we read off the desired elements of the anomalous
dimension matrix
\begin{equation}
  \gamma =  
 \begin{pmatrix}
 0 & 0 & 0 \\
 -2/3 & 0 & 0 \\
 \phantom{-}4/3 & 0 &  0
\end{pmatrix}  + {\mathcal O}(\alpha_s) \,,
\end{equation}
where the neglected higher-order terms describe the mixing of
``intrinsic-charm'' operators into themselves and of the 
Darwin term into itself, which are not explicitly needed
for the discussion in the body of the text.

\subsection{Dimension-7}

A similarly simplified analysis can be performed 
for the mixing of the dimension-7 ``intrinsic-charm'' operators
into the dimension-7 two-quark operator $m_c^4 \, \bar b_v \, \slashed{v} \, b_v$.
As before, defining 
\begin{align}
	{\cal O}_2 &= m_c^4 \, \bar{b}_v \, \slashed{v} \, b_v \,, \cr 
	{\cal O}_{T_3} &= (4\pi)^2 \, \mu^{2\epsilon} \, \frac{1}{3} \left( ( i v\cdot \partial \, \bar{b}_v \, \gamma_\mu P_L \, c) \, (\bar{c}\, \gamma^\mu P_L \, b_v) - (i v\cdot \partial \, \bar{b}_v \, \slashed{v} P_L \,  c) \, (\bar{c} \, \slashed{v} P_L \,  b_v) \right)\,, \cr
 	{\cal O}_{T_4} &= (4\pi)^2 \, \mu^{2\epsilon} \, \frac{1}{3} \left( (i\partial_\alpha \, \bar{b}_v \, \slashed{v} P_L \, c) \, (\bar{c}\, \gamma^\alpha P_L \, b_v) - (i v\cdot \partial \, \bar{b}_v \, \slashed{v} P_L \,  c) \, (\bar{c} \, \slashed{v} P_L \,  b_v) \right) \,, \cr 
 	{\cal O}_{T_5} &= (4\pi)^2 \, \mu^{2\epsilon} \,\frac{1}{3} \left( (i\partial_\alpha \, \bar{b}_v \, \gamma^\alpha P_L \, c) \, ( \bar{c}\, \slashed{v} P_L \, b_v) - (i v\cdot \partial \, \bar{b}_v \, \slashed{v} P_L \,  c) \, ( \bar{c} \, \slashed{v} P_L \,  b_v) \right) \,,  \cr 
	{\cal O}_{T_6} &= (4\pi)^2 \, \mu^{2\epsilon} \, \frac{1}{3} \left( 6 \, (i v\cdot \partial \, \bar{b}_v \, \slashed{v} P_L \,  c) \, (\bar{c} \, \slashed{v} P_L \,  b_v)- (i v\cdot \partial \, \bar{b}_v \, \gamma_\mu P_L \, c) \, (\bar{c}\, \gamma^\mu P_L \, b_v) \right) \cr 
	& \quad {} -(4\pi)^2 \, \mu^{2\epsilon} \, \frac{1}{3} \left(  (i \partial_\alpha \, \bar{b}_v \, \slashed{v} P_L \, c) \, (\bar{c}\, \gamma^\alpha P_L \, b_v) + (i \partial_\alpha \, \bar{b}_v \, \gamma^\alpha P_L \, c) \, (\bar{c}\, \slashed{v} P_L \, b_v) \right) \,, \cr 
	{\cal O}_{T_7} &= (4\pi)^2 \, \mu^{2\epsilon} \, \frac16 \, \epsilon^{\mu\nu\alpha\beta} \, v_\beta \, ( i\partial_\alpha \, \bar{b}_v \, \gamma_\nu P_L \,  c) \, (\bar{c} \, \gamma_\mu P_L \,  b_v)\,,
\end{align}
we calculate the contributions to the 2-parton matrix elements 
from the tadpole diagram in Fig.~\ref{figb}(a) as
\begin{align}
\langle b| {\cal O}_{T_3} |b \rangle^{(0)} 
     &= + \frac{1}{8} 
       \left(\frac{1}{\epsilon}+\ln \frac{\mu^2}{m_c^2} + \ldots \right) \langle b |{\cal O}_2|b\rangle_{\rm tree}
\,,
 \cr 
\langle b| {\cal O}_{T_4} |b \rangle^{(0)} = \langle b| {\cal O}_{T_5} |b \rangle^{(0)}
     &= - \frac{1}{8}  \left(\frac{1}{\epsilon}+\ln \frac{\mu^2}{m_c^2} + \ldots \right) \langle b |{\cal O}_2|b\rangle_{\rm tree}
\,,
 \cr 
\langle b| {\cal O}_{T_6} |b \rangle^{(0)} & = 0
\,,
\cr 
 \langle b| {\cal O}_{T_7} |b \rangle^{(0)}
     &= - \frac{1}{8}  \left(\frac{1}{\epsilon}+\ln \frac{\mu^2}{m_c^2} + \ldots \right) \langle b |{\cal O}_2|b\rangle_{\rm tree}
\,,
\end{align}
from which we read off the elements of the anomalous dimension matrix
entering (\ref{C2}).

\end{document}